# Angle-sensitive pixel design for wavefront sensing

Guoan Zheng*

*Electrical Engineering, California Institute of Technology, Pasadena, CA 91125, USA*
*Corresponding author: gazheng@caltech.edu*

Conventional image sensors are only responsive to the intensity variation of the incoming light wave. By encoding the wavefront information into the balanced detection scheme, we demonstrate an image sensor pixel design that is capable to detect both the local intensity and angular information simultaneously. With the full compatibility to the CMOS fabrication process, the proposed pixel design can benefit a variety of applications, including phase microscopy, lensless imaging and machine vision.

*OCIS Codes: 040.1240, 110.5200, 230. 4000, 050.1220,*

The concept of light field camera (or plenoptic camera) has received much attentions in recent years [1, 2]. Such a camera can capture both intensity and angular information of incoming light waves. Based on these two types of information, it is possible for the user to interactively change the focus, the view point and the perceived depth-of-field of the captured image upon digital post-processing. However, conventional CCD/CMOS image sensors can only capture a 2D intensity map of the incoming light wave; angular information is lost in the measurement process. To address this issue, several schemes have been developed to record both the intensity and the angular distribution of the light field, including the use of an array of conventional cameras [3], multiple masks in the optical path [4] and a microlens array [2]. These approaches recover the angular information based on the relative position between the external optical component and the imaging recording pixel array.

Recently, it is shown that, the measurement of the angular information can be integrated at the pixel level of a CMOS image sensor, termed light field imager [5-7]. Such a light field imager has been successfully demonstrated for synthetic refocusing, depth mapping, 3D localization and lensless imaging [5-7]. The key idea of this light field imager is to encode the angular information in the intensity measurement at pixel level. It employs a pair of diffraction gratings placed above photodiodes to achieve angular sensitivity. Upon illumination, the top grating generates diffraction patterns that have periodicity identical to the grating pitch (Talbot effect). The bottom grating is used to selectively transmit the diffracted light to the photodiode below. In such a pixel design, there are two ambiguity needed to be addressed: 1) the ambiguity between the local intensity and the local incident angle; 2) the intrinsic periodicity of the angular response. To resolve these two ambiguities, 8 pixels are needed to fully determine the local angular information in two dimensions.

In this letter, we present a simple angle-sensitive pixel (ASP) design based on the balanced detection scheme. We combined 4 conventional pixels to form one ASP group. The summation of pixel values represents the local light intensity and the difference of pixel values represents the local incident angle. This letter is structured as follows. We will first describe the principle of the proposed ASP. Next, we will report on our full wave simulation of the ASP design for two typical pixel structures. Finally, we will draw our conclusion at the end of this letter.

The proposed ASP design is shown in Fig. 1(a), where 4 conventional pixels share one top metal opening. The cross-section view in x-z plane is shown in Fig. 1(b1) and (b2). In the rest of the paper, we will focus our discussion on the angular response at x direction. The case for y direction can be treated in the same manner. As shown in Fig. 1(b1), upon normal incidence, readout of pixel 1 is exactly the same as pixel 2, and we refer this case as the balanced state. With a non-zero incident angle θ shown in Fig. 1(b2), the readout of pixel 1 is higher than that of pixel 2 and the incident angle can be recovered from the intensity difference of these two pixels.

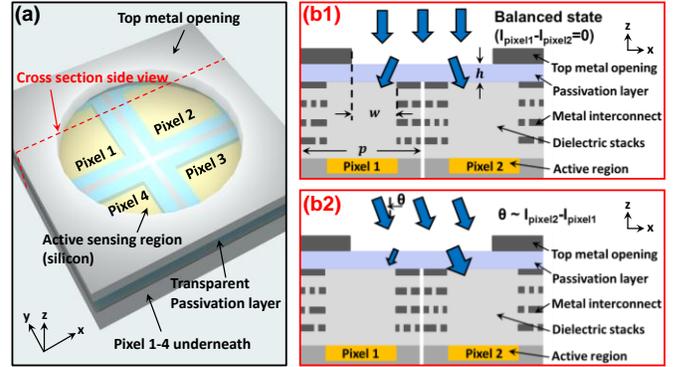

Fig. 1 (Color online) (a) The proposed ASP design. (b1) Balanced state of the ASP under normal incidence. (b2) The incident angle can be recovered based on the difference of the two pixel values.

In the large scale limit (i.e. the ray-optics limit), intensity readouts of pixel 1 and 2 can be expressed as

$$I_1 = I_0(w - h*tan\theta/n),\ I_2 = I_0(w + h*tan\theta/n) \quad (1)$$

, where $I_0$ is the local intensity of the incoming wave, $n$ is the refraction index of the passivation layer, $w$ is the opening size of individual pixels (denoted in Fig. 1(b1)), $h$ is the thickness of the top passivation layer. The local intensity of the light wave can be simply measured by pixel binding, i.e. $(I_2 + I_1)$. The incident angle in x direction can be recovered based on the following equation:

$$\theta = \arctan\left(\frac{n \cdot w}{h}\frac{I_2 - I_1}{I_2 + I_1}\right) = \arctan\left(\frac{n \cdot w}{h}\frac{\Delta I}{I_{total}}\right) \quad (2)$$

, where 'w/h' is a structure parameter to characterize the angular sensitivity of the ASP. A smaller 'w/h' promises a higher angular sensitivity, with the tradeoff in the total measurement range and the pixel crosstalk.

In a practical CMOS image sensor pixel design, the size of individual pixel is at the micrometer scale (For example, the pixel size of image sensor in most of mobile modules ranges from 1.1 to 3 micrometers). The light diffraction effect at this scale plays an important role in the angular response of the proposed ASP. Next, we will present our FDTD-based full-wave simulations for two types of ASP design, one for the front-side illuminated pixel structure and the other for the back-side illuminated structure.

Fig. 2 demonstrates an ASP design based on a 2.2 μm front-side illuminated pixel structure. In this design, the size of entire ASP is 4.4 μm; the top metal opening is 2.4 μm; the refraction index and the thickness of the passivation layer is 1.47 and 1.2 μm. To reduce the complexity of the simulation, we use perfect electric conductor as metal layers. The bottom part of the ASP is the silicon layer, where we define a 1.6 μm * 1 μm region (denoted by the dash line) as the active sensing area for power flow integration. Two types of simulation result are given in Fig. 2: the electric field (Fig. 2(a1) and (b1)) and the time averaged power flow (Fig. 2(a2) and (b2)). Fig. 2(a) demonstrates the case of the balanced state, corresponding to Fig. 1(b1). Fig. 2(b) demonstrates the case of 10 degree incidence, corresponding to Fig. 1(b2). In Fig. 1(b2), the power flow in the active region of pixel 2 is higher than that of pixel 1, and the ratio between $I_1$ and $I_2$ is determined to be 2.5 from this simulation.

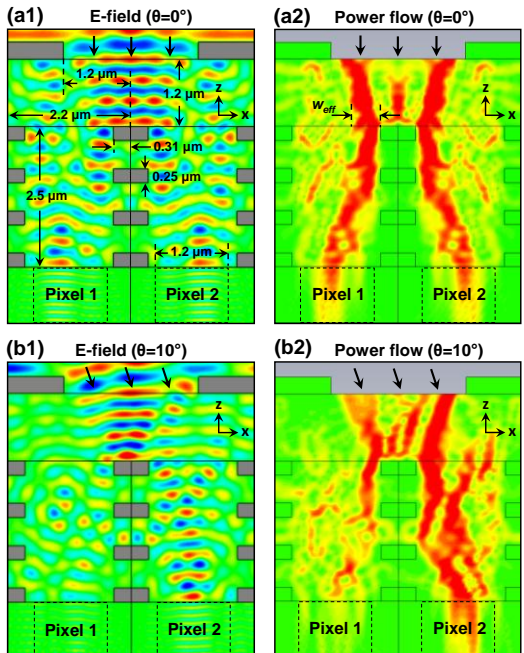

Fig. 2 (Color online) The FDTD simulation of the front-side illuminated ASP design with normal incidence (a) and 10 degree incidence (b). The wavelength is chosen to be 550 nm, the center of visible spectrum.

An important feature shown in Fig. 2 is the diffraction-based focusing effect of the top metal opening. In Fig. 2(a2), as the light wave passes through the top metal opening, the effective beam width decreases; in other words, the top metal opening acts an effective lens to focus the light wave into the center part of the ASP. In this regard, we can define an effective opening size $w_{eff}$ to correct for $w$ defined in Eq. (1). For Fig. 2(a2), the effective $w_{eff}$ is about 1.8 times smaller than the $w$.

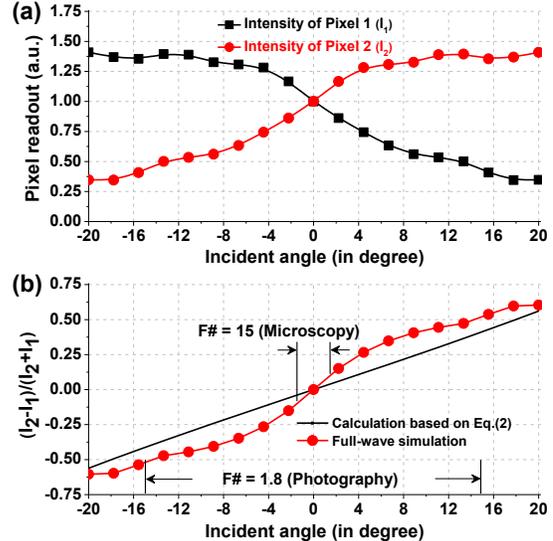

Fig. 3 (Color online) (a) Intensity readouts for pixel 1 and 2 with respect to the incident angle. (b) The angular response of the proposed ASP.

The angular response of the front-side illuminated ASP is shown in Fig. 3. We observe that $(I_2-I_1)/(I_2+I_1)$ is a monotonically increasing function with respect to the incident angle $\theta$. Therefore, there is no ambiguity for the angular response for different $\theta$s. In Fig. 3(b), we also compare the simulation result with the theoretical calculation based on Eq. (1). The effective opening size $w_{eff}$ is used to correct the focusing effect of the top metal opening. We can see that the overall trends of these two curves are in a good agreement with each other. However, there are also some discrepancies worth discussing. The simulated angular sensitivity is higher than that of Eq. (1) for small incident angles, and lower for large incident angles. Such discrepancies can be attributed to the uniform light ray assumption used in Eq. (1). Due to the focusing effect of the top metal opening, the actual power flow is stronger in the center of the ASP. Therefore, for small incident angles, the focused light wave enters into one of the individual pixels, resulting in a steeper slope of the angular response in Fig. 3(b). The lower angular sensitivity at larger incident angles can be explained in a similar manner.

The angular range of the simulation spans from -20 to +20 in Fig. 3. It covers most of applications in photography and microscopy. Based on the red curve in Fig. 3(b), we can also determine the minimum angular sensitivity of the proposed ASP. For photography application (with an f-number of 1.8), the minimum angular sensitivity locates at the largest incident angle (~15 degree). Assuming we have a 10 bit image sensor with 1023 intensity levels, the minimum angular variation we can detect is 0.47 mrad (in upper limit) in this case. For microscopy application, the angular range at the image plane is about -2 degree to +2 degree, corresponding to an f-number of 15. In this range, the minimum angular variation we can detect is 0.12 mrad, about 4 times better than the previous case. Such a high angular sensitivity at small angle range also perfectly fit into microscopy applications, where high accuracy is desired for quantitative analysis.

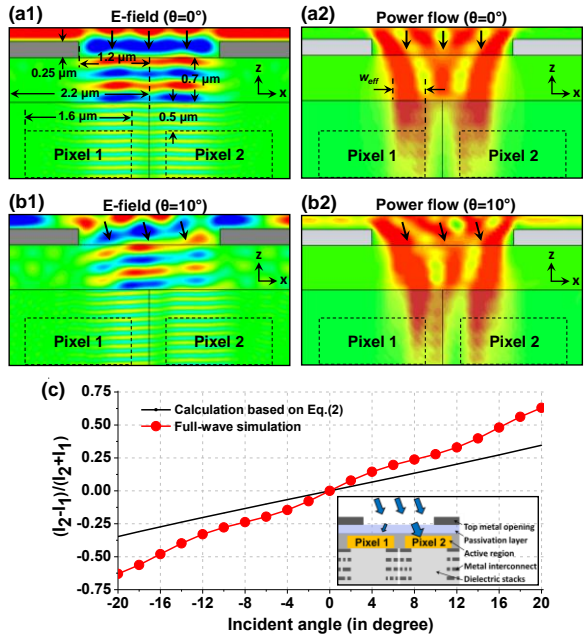

Fig. 4 (Color online) The FDTD simulation of the back-side illuminated ASP design with normal incidence (a) and 10 degree incidence (b). (c) The angular response of the proposed ASP.

Another trend of the CMOS image sensor development is the use of back-side illuminated pixel structure. Such a structure orients the wiring behind the photocathode layer by flipping the silicon wafer during manufacturing and then thinning its reverse side so that light can strike the photocathode layer without passing through the wiring layer. The proposed ASP can also be adapted to the back-side illuminated pixel structure. Fig. 4(a) and 4(b) show the FDTD simulation of the back-side illuminated ASP design. As shown in Fig. 4(a2), the effective opening size $w_{eff}$ is about 1.7 times smaller than $w$ in this case. In Fig. 4(c), we also compare the simulated angular response with Eq. (1). In such a back-side illuminated pixel structure, the active region locates at the bottom of the silicon layer. The effective layer height $h$ is larger than that of Eq. (1), and thus, the simulated angular response exhibits a steeper slope in Fig. 4(c). The minimum angular sensitivity of such an ASP design is about 0.3 mrad over the range of -20 degree to +20 degree.

To conclude, we have demonstrated a simple ASP design for both the front-side and back-side illuminated pixel structures. The proposed ASP employs the balanced detection scheme to measure the local intensity and the angular information simultaneously. The estimated angular sensitivity is about 0.1-0.4 mrad per intensity-level for a typical 10 bit CMOS image sensor. There are several advantages associated with the proposed ASP design:

1) No angular ambiguity. Unlike the diffraction-grating approach, the angular response of the proposed ASP is a monotonically increasing function with respect to the incident angle; therefore, there is no angular ambiguity for the proposed ASP.

2) High pixel density. In the proposed ASP, the recovery of the 2D angular information is based on intensity measurements of 4 conventional pixels. In other words, the ASP density is only 4 times less than that of a conventional CMOS imager, and it is generally higher than the microlens/pinhole-based wavefront sensor [8].

3) Full compatibility with the CMOS fabrication process. With only one extra metal layer on top, the proposed ASP design can be easily integrated in the conventional CMOS fabrication process. We can even directly modify an existing CMOS imager by post-fabrication processes [9, 10].

Finally, we note that, the concept of balanced detection scheme is not new. It has been demonstrated for wavefront detection at a relative large scale (~0.5 mm) [11]. However, we believe that the integration of such a scheme at the pixel level, especially in combination with the emerging back-side illuminated structure, enables a variety of new application possibilities in light field photography [2], phase microscopy [8], lensless imaging [6, 12] and machine vision. Some further studies of the proposed ASP, including the diffraction-based focusing effect, the optimal angular response and the optical confinement methods for adjacent ASPs [13], are highly desired in the future.